# Robustness and Stability of Enterprise Intranet Social Networks: The Impact of Moderators

Fronzetti Colladon, A., & Vagaggini, F.

# Robustness and Stability of Enterprise Intranet Social Networks: The Impact of Moderators

Fronzetti Colladon, A., & Vagaggini, F

**Abstract**

In this study, we tested the robustness of three communication networks extracted from the online forums included in the intranet platforms of three large companies. For each company we analyzed the communication among employees both in terms of network structure and content (language used). Over a period of eight months, we analyzed more than 52,000 messages posted by approximately 12,000 employees. Specifically, we tested the network robustness and the stability of a set of structural and semantic metrics, while applying several different node removal strategies. We removed the forum moderators, the spammers, the overly connected nodes and the nodes lying at the network periphery, also testing different combinations of these selections. Results indicate that removing spammers and very peripheral nodes can be a relatively low impact strategy in this context; accordingly, it could be used to clean the "noise" generated by these types of social actor and to reduce the computation complexity of the analysis. On the other hand, the removal of moderators seems to have a significant impact on the network connectivity and the shared content. The most affected variables are closeness centrality and contribution index. We also found that the removal of overly connected nodes can significantly change the network structure. Lastly, we compared the behavior of moderators with the other users, finding distinctive characteristics by which moderators can be identified when their list is unknown. Our findings can help online community managers to understand the role of moderators within intranet forums and can be useful for social network analysts who are interested in evaluating the effects of graph simplification techniques.

**Keywords:** enterprise intranet; online forum networks; moderators; robustness; stability; network analysis.

# 1. Introduction

Many complex systems can be described as networks, in which the constituent components are represented by vertices and their connections represent the relationship existing between them. A fundamental issue concerning complex networks is the robustness of the overall system to the failure of its constituent parts. Network robustness is defined as the ability to retain one or more specific properties under perturbation of its structure, e.g. after node or edge removal (Albert, Jeong, & Barabasi, 2000; Barabasi, 2016). Robust networks are more tolerant to random failures and less vulnerable to intentional attacks against nodes and edges (Albert et al., 2000; Barrat, Bathelemy, & Vespignani, 2008). The concept of robustness is often linked to the study of network resilience, which is partially different as it usually implies the additional analysis of the network evolution over time, after potentially harmful events. Network resilience is defined as "the ability of a system to adapt to errors or intentional attacks, by changing its mode of operation without losing its ability to function" (Barabasi, 2016, p. 303; Gao, Barzel, & Barabási, 2016). An understanding of vulnerabilities of complex networks is fundamental to the design of robust systems (Estrada, 2006), as well as to the definition of strategies for the management of their social impact (Holme, 2004).

Following this approach, we investigated the concepts of network robustness and stability in the context of enterprise forum networks, used by employees to communicate and share knowledge. We studied the interactions taking place among the employees of three large companies, using the forums provided on the company intranets. Analyzing the interaction patterns in these three intranets, over an eight-month period, we were able to map three different networks, where employees accounts (network nodes) are linked together if they interacted in the same forum post, for instance answering to each other's comments.

## 1.1. Research Objective

Considering the three enterprise intranets in our case study, we specifically analyzed the network robustness and the stability of well-known structural metrics, while simultaneously removing nodes based on their category (e.g. Moderators or Spammers), or on their structural properties, such as their degree centrality. In addition, we investigated the changes produced in the shared content, in terms of language sentiment, emotionality and complexity. Specifically, we focused our attention on forum moderators - who managed the forum content moderating discussions and posting messages which conveyed the institutional view of the company management -  and on some removal strategies aimed at reducing the "noise" generated by other specific categories of nodes:  the  spammers, identified as those social actors who post mostly undesired and irrelevant messages; the loosely connected nodes, lying at the network periphery; the overly connected nodes, as possible alternative spammers identification.

Our research has two main objectives: first, we want to help network analysts understand which users can be removed from intranet forum networks without significantly harming the general connectivity and the sentiment of the language used, i.e. keeping structural and semantic variables stable. Simplifying the network graph, before carrying a social network analysis, is indeed often useful to reduce the computational complexity of some algorithms and to clean the signal from possible disturbances introduced, for instance, by irrelevant messages posted by spammers. Secondly, we try to help community managers to better understand the role of moderators and their impact on the internal communication processes. Understanding the distinctive behaviors of moderators and how they influence the network connectivity can be very useful for managers who want to foster and optimize the social interaction among employees.

**2.  Literature Review**

The properties of robustness and resilience of complex networks to attacks or failures, on nodes or edges, has been widely investigated over many disciplines (Gao, Liu, Li, & Havlin, 2015), such as the World Wide Web (Albert et al., 2000; Boldi, Rosa, & Vigna, 2013; Broder et al., 2000), social and collaborative networks (Baek, Meroni, & Manzini, 2015), enterprise communication systems and e-mail networks (Aedo, Díaz, Carroll, Convertino, & Rosson, 2010; Wang, Gao, & Ip, 2010), and other disciplines like supply chain management (Tang, Jing, He, & Stanley, 2016), socio-ecological systems (Crespo, Suire, & Vicente, 2014), or biological networks (Aerts, Fias, Caeyenberghs, & Marinazzo, 2016).

There are several different ways by which vertices and edges can be removed and networks can show varying degrees of robustness to these strategies (Braunstein, Dall'Asta, Semerjian, & Zdeborová, 2016). We distinguish between network failures, which are typically attacks without a prior knowledge of the network structure or due to unexpected system errors, and targeted attacks which are usually meant to maximize the damage to the network connectivity and functioning. Nodes and edges can be removed simultaneously or sequentially (Holme, Kim, Yoon, & Han, 2002; Ventresca & Aleman, 2015). In a sequential attack, nodes or edges are progressively removed, each time observing the effect on the overall system, before deciding the next target. By contrast, a simultaneous attack is implemented at the same time against a previously selected subset of nodes or edges.

Investigations about robustness and resilience of real world networks have been centered on the analysis of several parameters such as connectivity, network distances, stability of centrality measures, average path lengths and clustering coefficients (Cohen & Havlin, 2010; Larhlimi, Blachon, Selbig, & Nikoloski, 2011). If an attacker desires to severely harm the functionality of a network, an intuitive strategy would be to target more critical social actors. Without prior information about the phenomenon represented by the network, only its structure can be examined, so the best strategies will target important nodes, such as those with a high degree or betweenness centrality (Borgatti, Everett, & Johnson, 2013).

Albert et al., (2000) sparked considerable interest in network robustness. They analyzed the impact of random failures and targeted attacks on two real-world graphs representing the topology of the Internet and a subset of the World Wide Web pages; both these graphs showed a power-law degree distribution (Broder et al., 2000). Studying the trend of the mean distance between vertices as a function of the number of removed nodes, they concluded that scale-free networks are extremely tolerant to random failures, but severely affected by targeted attacks. Broder et al. (2000) came independently to a similar conclusion studying a larger subset of the World Wide Web. Authors found that all nodes with a degree greater than five should be removed to significantly harm the graph connectivity. Accordingly, they maintained that the network was robust against targeted attacks, which seems to contrast the work of Albert et al., (2000). However, even if removing nodes with degrees greater than five seems to be a massive attack on the graph, there is no conflict between these results "because of the highly skewed degree distribution of the Web, the fraction of vertices with degree greater than five is only a small fraction of all vertices" (Newman, 2003, p. 16). The context of the World Wide Web is somewhat close to our study of intranet social networks, where we also observe a degree distribution similar to a power law.

    A particularly thorough study of the robustness of both real-world and model networks was conducted by Holme et al. (2002) who investigated vertex and edge removal considering both inverse geodesic length and the size of the largest connected subgraph. They considered four removal strategies, to be implemented over time: initial degree, initial betweenness, recalculated degree, and recalculated betweenness. Other scholars considered the effects of nodes removal on the size of the giant component, i.e. the largest significant connected component of the graph. If the size of the largest component is small, with respect to the size of the system, it is often reasonable to assume that the network will be unable to function properly. Therefore, the network is considered connected if the giant component still exists after the deletion of several vertices, so that the network can continue its main operations, even in case of temporal node

malfunctioning (Chen & Cheng, 2015). Similar results were obtained in the experiment of Broder et al. (2000), who looked at the behavior of the World Wide Web graph as vertices were removed in order of decreasing degree. Braunstein et al. (2016) explored the network dismantling problem, proposing a new algorithm to determine the minimal sets of nodes that, if removed from a large graph, would determine a fragmentation into smaller connected components. Schneider et al. (2011) also looked at the size of the giant component during malicious attacks and used their results to propose a method to efficiently mitigate the risk. They tested their model on the European power grid and on the Internet, as well as on other complex networks models. Iyer and colleagues (2013) investigated the effect of removal schemes based on degree, betweenness, closeness, and eigenvector centrality, while varying the network clustering coefficient and the level of assortativity for exponential and scale-free networks. They used a vulnerability measure, again based on the giant component, showing that sequential attacks can be more effective than simultaneous ones.

Shchieber et al. (2016) proposed a dynamical analysis of network robustness, based on sequential failures, considering the evolution of the network diameter and the dissimilarities between sequential topologies. Another measure related to the network's robustness and resilience is the assortative coefficient, i.e., the degree-degree correlation level. Rubinov and Sporns (2010), showed how networks with a positive assortative coefficient are expected to have a resilient core of interconnected nodes with high degrees, whereas disassortative networks present widely distributed hubs and are consequently more vulnerable. Recently, other scholars tried to develop new robustness estimation techniques, considering multiple measures together and trying to minimize their computation time (Wandelt, Sun, Zanin, & Havlin, 2017).

Other studies were more concerned with attack strategies. Trajanovski et al. (2013), for instance, examined network robustness considering both random node failure and simultaneous attack based on five network centrality measures. Their results indicate that many centrality measures may produce similar results, but degree or eigenvector centrality may suffice to evaluate worst-case behaviors. They also found that increasing assortativity is beneficial when protecting against targeted attacks, but is not the best approach to protect against random failures. Similarly, Yehezkel and Cohen (2012) analyzed the stability of random scale-free networks to degree-dependent attacks, proposing an optimal defense strategy based on the addition of links to average degree nodes. In order to increase realism in the attacks, Gallos et al. (2006) studied robustness of scale-free networks to a number of attack strategies, while varying the amount of information available to the attacker. Xiao et al. (2008) observed that the intentional attack is the most effective strategy to disrupt a whole system, when the adversary has a full knowledge of the network structure. Booker (2012) studied how targeting strategies can be affected by erroneous data about the network structure. His findings suggest that even removing a small amount of appropriately targeted nodes can have a devastating effect and that the best protection is often to hide critical nodes. More recently, Gao et al. (2016) developed a set of analytical tools meant to identify the natural control and state parameters of a multi-dimensional complex system, helping in predicting the system's resilience.

Lastly, some scholars focused their attention on the stability of structural metrics, while removing nodes from the original network. Consteinbader and Valente (2003) examined the stability of eleven centrality measures for several network structures, after bootstrap sampling. With their simulations, authors showed that there are significant differences, with some measures being more stable than others.

The above-mentioned literature seems to suggest that targeted nodes removals – mostly based on degree centrality - are often more harmful than random failures. In our research, we analyzed real enterprise intranet social networks (without the possibility of an actual nodes removal, which would have implied the blocking of specific user accounts). For this reason, we focused on the study of network robustness and on the stability of structural and semantic metrics, testing targeted attack strategies, based on simultaneous nodes removals. Accordingly, we considered the following research questions:

RQ1a: Which nodes (user accounts) can be removed from online forum networks without significantly harming their connectivity and keeping structural and interaction metrics stable?

RQ1b: While removing different sets of nodes, which are the most affected structural and semantic variables?

## 2.1. The Role of Moderators and Spammers in Forum Networks

Moderators are social actors who act as guardians of the forum content; they can moderate discussions, post messages to enforce rules of conduct, or guide users through a more appropriate use of the forum and help them to find an answer to their questions (e.g., Brace-Govan, 2003). Moderators are also important as they bring in their expertise about the forum topics (Huh, Yetisgen-Yildiz, & Pratt, 2013); they organize posts, create and edit forum content, and can shape the overall tone of the communication (Frith, 2014); they often act as discussion leaders, facilitating the discourse and keeping down "flames" (Berge & Collins, 2000).

In our case studies, moderators are internal employees specifically engaged by company management. They have the assignment to interact with the other employees, sharing news and answering questions and comments related to the company and its events. They are also engaged with the purpose of signaling to the management potential cases of employees who are openly disappointed with company policies and behaviors. Only extreme cases are reported – where employees are, for instance, publicly insulting the management or their choices (such cases were sporadic to non-existent in the networks we analyzed). The aim is not to repress the voices of unhappy employees, but to activate internal processes which could prevent and reduce their disengagement (Gloor, Fronzetti Colladon, Grippa, & Giacomelli, 2017). Given their important role, we expect that the removal of moderators could have significant impact on the structural properties of a forum network, as well as on the shared content.

The role of moderators is generally formalized: they are often nominated by the forum administrators and their role is usually known by the forum users. Spammers, on the other hand, are more frequently hidden within communication networks. They can be identified as those social actors who post a relatively high number of undesired and irrelevant messages (Heydari, Tavakoli, Salim, & Heydari, 2015; Stringhini, Kruegel, & Vigna, 2010). While moderators are usually clearly identified on the forums, the spammer nodes need to be identified through manual detection or using automated procedures - such as a machine learning approach (Bouguila & Amayri, 2009; Chakraborty, Pal, Pramanik, & Ravindranth Chowdary, 2016) - when a manual detection can be time consuming. In our research, we considered as "spammers" those social actors who typically spam the forum with undesired and mostly unread messages, such as advertising, or false news. We support the idea that an efficient identification of spammers should consider both the content of messages and the social behavior of users (Gayo-Avello, 2013; Yu, Chen, Jiang, Fu, & Qin, 2017); specifically, we considered spammers those users who met at least two of the following conditions:

a) Posted a very high number of messages;
b) Received zero or few answers by other users (not identified as spammers as well);
c) Messages content is manually classified as spam, or as a general message directed to all the employees (this includes, for instance, advertising messages, but excludes relevant HR communications, which are often answered and commented).

The manual detection of spammers, which we implemented in our case study, can be useful to overcome one limitation of simulated approaches, where it can be more difficult to differentiate between spammers and nodes that are just overly connected. However, trained machine learning approaches might come to help, in real cases, to automatically classify the content of the messages posted by each user (Crawford, Khoshgoftaar, Prusa, Richter, & Al Najada, 2015).

While the social behavior of spammers has been more widely investigated (Chakraborty et al., 2016; Zheng, Zeng, Chen, Yu, & Rong, 2015), the role of moderators remains less explored.

For this reason, we decided to make the study of the distinctive characteristics of these social actors the focus of our second research question.

RQ2: Can moderators be identified by a different use of language and by distinctive social behaviors?

## 3. Methodology

In this study, we analyzed three real world enterprise communication networks, extracted from the forums available on their intranet platforms. The three companies all operate in the same business sector and have a comparable medium-large size; as per agreed privacy arrangements, we are prohibited from revealing sensitive data, therefore the three forum networks will be identified as Network A, Network B and Network C. In each network user accounts are represented as nodes, existing if a specific employee posted some news or a comment during the study period; nodes are connected by directed links, if users answered one another. The first network was extracted from an enterprise intranet social network in its launch phase. We collected about 13,000 messages posted by more than 3,200 employees over a period of 8 months. Network B and Network C were constructed collecting over 24,000 and 15,000 messages respectively, from more than 5,200 and 3,700 employees, during an 8-month period. Differently from Network A, Networks B and C represent forums at a maturity stage, some years after launch. Apart from these differences, the forum structures and interfaces were very similar for the three companies. The relative number of spammers, moderators and peripheral nodes was also similar for the three intranet social networks (see Table 1a, 1b, 1c).

All the three networks show a degree distribution which can be approximated to a scale-free network, since they present a larger number of vertices with few links and a smaller number of hubs with many connections.

With the aim of observing the effects produced both on individual and whole network metrics, we tested several different removal strategies, each of which had a different objective. Specifically we removed:

a) the Moderators, to clean the network from the contributions of people who fulfill an institutional role (conveying the point of view of company management and being on average more active, because of their work assignment). The list of Moderators was provided to us by the three companies;

b) the Spammers, to clean the network from undesired and irrelevant messages;

c) the combination of moderators and spammers;

d) the nodes in the top first, fifth and tenth percentile of the network degree distribution, to see the effects produced by the deletion of nodes with a high degree centrality. In addition, we implemented this strategy for comparative purposes, as it is widely used in the robustness literature. Such an approach can also be useful when the spammers can only be recognized as the most connected nodes;

e) the nodes at the bottom of the network degree distribution, i.e. isolated nodes and those connected by a single link, for their negligible role on large networks;

f) the combination of the nodes in the top first and bottom selections;

g) the combination of bottom nodes and moderators;

h) the combination of bottom nodes and spammers.

## 3.1. Variables Description

We represented the three forum networks as directed graphs $G = (V, A)$, where $V = \{v_1, v_2, ... v_n\}$ is the set of nodes, and $A$ is the set of directed arcs ($a_{ij}$).

The effects of the removal strategies, detailed in the previous section, were analyzed considering their impact on whole network metrics, such as diameter or average distance among reachable pairs, and on individual metrics, such as degree or betweenness centrality, or sentiment, emotionality and complexity of the language used (Brönnimann, 2014). Each specific measure analyzed is better described in the following.

### 3.2. Whole network metrics

*Average Distance Among Reachable Pairs (ADARP).* This measure calculates the average length of the shortest paths between all pairs of nodes in the graph that can reach one another, i.e. when the shortest path has a length lower than infinite (De Nooy, Mrvar, & Batagelj, 2012).

*Diameter (D).* The diameter of a network is the largest geodesic distance in the connected network.

*Clustering Coefficient (CC).* This measure is calculated counting the number of closed triplets of nodes and dividing it by the total number of connected triplets in the network (De Nooy et al., 2012). It provides information on how nodes tends to cluster together, forming highly connected groups, with a lower number of connections between different clusters.

*Average Degree (AD).* This measure is equal to the average number of ties that originate or terminate at each node.

### 3.3. Node-level metrics

*Degree Centrality*. This variable is calculated as the number of links adjacent to a node. It gives us information on how many other actors can be directly reached by a social actor (Borgatti et al., 2013).

*Closeness Centrality.* This variable is defined as the inverse of the distance of a node from all others in a network, considering the shortest paths that connects each pair of nodes, and it measures the mean distance from a vertex to all the others. Closeness centrality can be used as a proxy for the speed by which a social actor can reach his/her peers (Borgatti et al., 2013).

*Betweenness Centrality.* This measures the extent to which a vertex lies in the paths between other vertices and it is equal to the number of times that a node is included in the shortest paths that connect every other pair of nodes (Borgatti et al., 2013). Vertices with high values of betweenness centrality may have considerable influence within a network by virtue of their control over information passing among others.

*Betweenness Oscillations.* This measure counts the oscillations in betweenness centrality for a node $v_i$, within a specific time interval. The more frequently a node changes its position, reaching local maxima and minima in its betweenness centrality scores, the higher the oscillations. We have no oscillations if a vertex keeps its position with respect to others. This measure has been introduced by Kidane and Gloor, (2007) to predict group creativity, and it has also been used to forecast business success in collaborative innovation (Davis & Eisenhardt, 2011) and innovative potential of startups (Allen, Gloor, Fronzetti Colladon, Woerner, & Raz, 2016).

*Activity.* In the forum network this counts the number of messages posted by a specific actor.

*Contribution Index.* This measure expresses the balance between messages sent and received by a social actor. The contribution index of the node $v_i$ is calculated dividing the difference between messages posted and received by the sum of these two measures. Contribution Index varies in the range [1,-1], where a contribution index of -1 identifies inactive actors that only receive messages, while 1 identifies people who only post messages without receiving any answer (Gloor, Laubacher, Dynes, & Zhao, 2003). Spammer nodes usually show high values of contribution index and, in our dataset, they all have a contribution index bigger than 0.7.

*Average Response Time (ART)*. This variable measures the average time taken by an employee to answer to forum posts directed to him/her (*Ego ART*); we also considered the symmetrical measure – the *Alter ART* – which represents the time taken by the other users to answer to his/her messages (Gloor, Almozlino, Inbar, Lo, & Provost, 2014).

*Nudge*s. This measure counts the average number of pings, i.e. nudges required before a social actor answers to a message (Gloor et al., 2014). Nudges can be subsequent messages sent to a user which has not yet answered. Similarly to the Average Response Time, this measure can be referred to the nudges sent by the node $v_i$ (*Ego Nudges*), or to the nudges sent from the other nodes to $v_i$ (*Alter Nudges*).

### 3.4. Semantic variables

*Sentiment.* It measures the sentiment of the language used in message bodies and it varies in the range [0,1], where a positive sentiment is characterized by values near to 1, values around 0.5 express a neutral sentiment, while lower values represent a negative sentiment. This variable was calculated using the multi-lingual classifier (Brönnimann, 2014) included in the software Condor (Gloor, 2017).

*Emotionality.* Calculated as the standard deviation of sentiment (Brönnimann, 2014), this variable expresses the level of emotionality in the language used. A high level of emotionality is a symptom of a vivid debate, with strong alternation of messages with positive and negative sentiment.

*Complexity*. This represents the average complexity of the vocabulary used and it is calculated as the likelihood distribution of words within a text – i.e., the probability that each word of a dictionary appears in the text (Brönnimann, 2014). We consider a word as more complex when it rarely appears in the context analyzed, and not when it is just a rare word.

We used the social network analysis software Condor and the software Pajek (De Nooy et al., 2012) to calculate the above measures. For our experiment, we selected variables which are widely used and which proved their important role in past studies (Gloor, Fronzetti Colladon, Giacomelli, Saran, & Grippa, 2017; Kidane & Gloor, 2007). A different selection of variables would always be possible, searching for other structural or semantic metrics. To replicate the process, the selection of variables could be adjusted by network analysts depending on the contexts analyzed, before choosing a graph simplification technique.

## 4. Results

The node removal strategies detailed in Section 3 were applied to the three enterprise forum networks. Table 1a, Table 1b and Table 1c show the effects produced by each strategy on the clustering coefficient, the diameter, the average degree and the average distance among reachable pairs. The tables also show the relative proportions of removed nodes with respect to each strategy, which are similar for the three networks.

| Forum Network A | Number of Nodes Removed | Percentage of Nodes Removed | ADARP | CC | AD | D |
|---|---|---|---|---|---|---|
| Full Network | - | - | 3.301 | 0.586 | 4.275 | 9 |
| Removed Top 1st Percentile | 32 | 1.0% | 7.582 | 0.042 | 1.707 | 45 |
| Removed Top 1st and Bottom Nodes | 157 | 4.8% | 7.547 | 0.039 | 1.720 | 45 |
| Removed Top 5th Percentile | 162 | 5.0% | 13.148 | 0.020 | 1.273 | 52 |

| Forum Network A | Number of Nodes Removed | Percentage of Nodes Removed | ADARP | CC | AD | D |
|---|---|---|---|---|---|---|
| Removed Top 10th Percentile | 325 | 10.0% | 27.993 | 0.013 | 1.084 | 79 |
| Removed Bottom Nodes | 125 | 3.8% | 3.365 | 0.581 | 4.345 | 8 |
| Removed Moderators | 83 | 2.6% | 6.306 | 0.127 | 2.108 | 64 |
| Removed Moderators and Bottom Nodes | 200 | 6.2% | 6.295 | 0.126 | 2.140 | 64 |
| Removed Spammers | 18 | 0.6% | 3.302 | 0.586 | 4.270 | 9 |
| Removed Spammers and Bottom Nodes | 143 | 4.4% | 3.265 | 0.582 | 4.340 | 8 |
| Removed Moderators and Spammers | 101 | 3.1% | 6.333 | 0.127 | 2.105 | 64 |

**Table 1a.** Whole network metrics in Forum Network A.

| Forum Network B | Number of Nodes Removed | Percentage of Nodes Removed | ADARP | CC | AD | D |
|---|---|---|---|---|---|---|
| Full Network | - | - | 3.306 | 0.569 | 4.794 | 8 |
| Removed Top 1st Percentile | 52 | 1.0% | 6.575 | 0.060 | 1.917 | 24 |
| Removed Top 1st and Bottom Nodes | 264 | 5.0% | 6.538 | 0.054 | 1.943 | 23 |
| Removed Top 5th Percentile | 262 | 5.0% | 10.867 | 0.028 | 1.407 | 40 |
| Removed Top 10th Percentile | 524 | 10.0% | 18.454 | 0.016 | 1.143 | 50 |
| Removed Bottom Nodes | 212 | 4.0% | 3.272 | 0.562 | 4.889 | 6 |
| Removed Moderators | 116 | 2.2% | 5.340 | 0.105 | 2.203 | 19 |
| Removed Moderators and Bottom Nodes | 323 | 6.2% | 5.301 | 0.102 | 2.239 | 19 |
| Removed Spammers | 26 | 0.5% | 3.306 | 0.569 | 4.791 | 8 |
| Removed Spammers and Bottom Nodes | 238 | 4.5% | 3.327 | 0.562 | 4.885 | 6 |
| Removed Moderators and Spammers | 142 | 2.7% | 5.338 | 0.105 | 2.200 | 19 |

**Table 1b.** Whole network metrics in Forum Network B.

| Forum Network C | Number of Nodes Removed | Percentage of Nodes Removed | ADARP | CC | AD | D |
|---|---|---|---|---|---|---|
| Full Network | - | - | 3.175 | 0.582 | 4.545 | 7 |
| Removed Top 1st Percentile | 37 | 1.0% | 6.675 | 0.071 | 1.830 | 25 |
| Removed Top 1st and Bottom Nodes | 135 | 3.6% | 6.669 | 0.068 | 1.844 | 25 |
| Removed Top 5th Percentile | 186 | 5.0% | 12.052 | 0.022 | 1.317 | 42 |
| Removed Top 10th Percentile | 373 | 10.0% | 25.868 | 0.019 | 1.113 | 86 |
| Removed Bottom Nodes | 98 | 2.6% | 3.151 | 0.575 | 4.597 | 7 |
| Removed Moderators | 98 | 2.6% | 4.615 | 0.226 | 2.579 | 24 |
| Removed Moderators and Bottom Nodes | 191 | 5.1% | 4.596 | 0.225 | 2.611 | 24 |
| Removed Spammers | 16 | 0.4% | 3.175 | 0.582 | 4.544 | 7 |
| Removed Spammers and Bottom Nodes | 114 | 3.1% | 3.152 | 0.576 | 4.596 | 7 |
| Removed Moderators and Spammers | 114 | 3.1% | 4.616 | 0.227 | 2.579 | 24 |

**Table 1c.** Whole network metrics in Forum Network C.

All the three networks exhibit a relatively small proportion of spammers; this can be partially explained by the fact that each user in the forums had to show his/her real name and the forums were not open to external guests or anonymous users. On the other hand, the values for the number of nodes with very few connections (network periphery) are higher. Nonetheless, the removal of the bottom nodes does not seem to greatly affect any of the four metrics considered, with just a slight reduction in the diameter for Network A and B and an expected small increase in the average degree. In general, when the removal of the bottom nodes is applied in combination with another removal strategy, it does not have a significant effect on the whole network metrics, compared to the application of the other strategy alone. Similarly, the removal of spammers show an even lower impact on whole network metrics for all the three forum networks (also due to their very small number in our case study, less than 1% in all networks).

On the other hand, the removal of moderators seems to have a strong impact on the whole network metrics – this effect is even stronger for the intranet social network in its launch phase (Network A), suggesting that the role of moderators is more vital when a new communication platform is being set up. However, we can also see a significant variation in all metrics for networks B and C. For all three networks the removal of moderators tends to halve the values of average degree, sharply increase ADARP and diameter, and significantly reduce the clustering coefficient. Finally, the simultaneous removal of spammers and moderators shows almost the same effect as when only moderators are removed. Results produced by the removal of nodes in the top 1st, 5th and 10th percentile of the degree distributions seems to be in accordance with previous studies conducted on the World Wide Web where it was found that a targeted removal of the highest degree vertices causes the increase of the mean vertex-vertex distance (Albert et al., 2000; Broder et al., 2000). In all three networks, removal of the top nodes causes a deep change in the connection patterns linking the social actors, with effects similar to the removal of moderators, but greatly amplified, especially when removing nodes in top 5th and 10th percentile of the degree distribution. Nonetheless, we can notice that for Network A, the removal of moderators has a much higher impact on the Diameter compared to the removal of overly-connected nodes.

Looking at these metrics, it seems reasonable for an analyst to remove the spammers and the peripheral nodes from an enterprise communication network; by contrast, removing moderators can strongly harm the network connectivity and affect the results of the analysis.

Tables 2a, Table 2b and Table 2c show the effects on the stability of individual level metrics and semantic variables. Each measure is presented as the correlation between the values of the metrics after the application of a removal strategy and the correspondent value in the original network.

|  | Removed Nodes | | | | | | | | | |
|---|---|---|---|---|---|---|---|---|---|---|
| **FORUM NETWORK A** (Pearson's Correlation Coefficients) | **Top 1st Percentile** | **Top 1st Perc. and Bottom** | **Top 5th Percentile** | **Top 10th Percentile** | **Bottom Nodes** | **Moderator Nodes** | **Moderators and Bottom Nodes** | **Spammer Nodes** | **Spammer and Bottom Nodes** | **Spammer and Mod. Nodes** |
| **Alter ART** | 0,9815 | 0,9795 | 0,9916 | 1,0000 | 0,9984 | 0,9999 | 0,9977 | 1,0000 | 0,9984 | 0,9999 |
| **Ego ART** | 0,9796 | 0,9794 | 0,9589 | 1,0000 | 1,0000 | 0,9980 | 0,9979 | 1,0000 | 1,0000 | 0,9980 |
| **Alter Nudges** | 0,9647 | 0,9149 | 1,0000 | 1,0000 | 0,9782 | 1,0000 | 0,9734 | 1,0000 | 0,9782 | 1,0000 |
| **Ego Nudges** | 0,9191 | 0,8977 | 1,0000 | 1,0000 | 1,0000 | 0,9230 | 0,9159 | 1,0000 | 1,0000 | 0,9230 |
| **Activity** | 0,9504 | 0,9496 | 0,7472 | 0,5839 | 0,9997 | 0,9738 | 0,9733 | 1,0000 | 0,9997 | 0,9739 |
| **Contribution Index** | 0,5845 | 0,5162 | 0,3619 | 0,2668 | 0,9806 | 0,6139 | 0,5417 | 0,9882 | 0,9721 | 0,5981 |
| **Betweenness Centrality** | 0,7041 | 0,6957 | 0,4921 | 0,2156 | 0,9997 | 0,9113 | 0,9072 | 1,0000 | 0,9997 | 0,9101 |
| **Betw. Cent. Oscillations** | 0,9353 | 0,9266 | 0,7805 | 0,6527 | 0,9898 | 0,9449 | 0,9384 | 0,9994 | 0,9894 | 0,9435 |
| **Closeness Centrality** | 0,2437 | 0,0735 | 0,1262 | 0,0535 | 0,9916 | 0,2781 | 0,1872 | 0,9576 | 0,9916 | 0,2787 |
| **Degree Centrality** | 0,9502 | 0,9531 | 0,8367 | 0,6901 | 0,9999 | 0,9938 | 0,9940 | 1,0000 | 0,9999 | 0,9938 |
| **Sentiment** | 0,8754 | 0,8468 | 0,8740 | 0,2203 | 0,9982 | 0,9077 | 0,9046 | 0,9983 | 0,9982 | 0,9085 |
| **Emotionality** | 0,8494 | 0,8242 | 0,8741 | 0,7741 | 0,9990 | 0,9168 | 0,9104 | 0,9987 | 0,9994 | 0,9161 |
| **Complexity** | 0,8357 | 0,8000 | 0,8498 | 0,9314 | 0,9982 | 0,8967 | 0,8889 | 0,9983 | 0,9981 | 0,8977 |

*Note.* All correlations are significant with p < .01.

**Table 2a.** Stability of structural and semantic variables in Network A.

|  | Removed Nodes | | | | | | | | | |
|---|---|---|---|---|---|---|---|---|---|---|
| **FORUM NETWORK B** (Pearson's Correlation Coefficients) | **Top 1st Percentile** | **Top 1st Perc. and Bottom** | **Top 5th Percentile** | **Top 10th Percentile** | **Bottom Nodes** | **Moderator Nodes** | **Moderators and Bottom Nodes** | **Spammer Nodes** | **Spammer and Bottom Nodes** | **Spammer and Mod. Nodes** |
| **Alter ART** | 0,9629 | 0,9611 | 0,9818 | 1,0000 | 0,9990 | 0,9769 | 0,9768 | 1,0000 | 0,9990 | 0,9769 |
| **Ego ART** | 0,9631 | 0,9629 | 0,9899 | 0,9879 | 1,0000 | 0,9514 | 0,9514 | 1,0000 | 1,0000 | 0,9514 |
| **Alter Nudges** | 0,9662 | 0,9295 | 1,0000 | 1,0000 | 0,9820 | 1,0000 | 1,0000 | 1,0000 | 0,9820 | 1,0000 |
| **Ego Nudges** | 0,5774 | 0,5368 | 0,7030 | 1,0000 | 1,0000 | 0,4437 | 0,4436 | 1,0000 | 1,0000 | 0,4437 |
| **Activity** | 0,9492 | 0,9484 | 0,6297 | 0,6297 | 0,9998 | 0,9846 | 0,9846 | 1,0000 | 0,9998 | 0,9845 |
| **Contribution Index** | 0,5793 | 0,5232 | 0,3340 | 0,2516 | 0,9872 | 0,6851 | 0,6361 | 0,9969 | 0,9836 | 0,6817 |
| **Betweenness Centrality** | 0,8113 | 0,8069 | 0,5297 | 0,2986 | 0,9998 | 0,9676 | 0,9670 | 1,0000 | 0,9998 | 0,9677 |
| **Betw. Cent. Oscillations** | 0,9244 | 0,9155 | 0,7934 | 0,6528 | 0,9938 | 0,9414 | 0,9354 | 0,9992 | 0,9930 | 0,9409 |
| **Closeness Centrality** | 0,3964 | 0,3753 | 0,1821 | 0,0992 | 0,9887 | 0,4909 | 0,5034 | 1,0000 | 0,9886 | 0,4886 |
| **Degree Centrality** | 0,9521 | 0,9544 | 0,8615 | 0,7214 | 0,9999 | 0,9900 | 0,9902 | 1,0000 | 0,9999 | 0,9900 |
| **Sentiment** | 0,8339 | 0,8170 | 0,8590 | 0,8995 | 0,9986 | 0,8564 | 0,8545 | 0,9989 | 0,9991 | 0,8573 |
| **Emotionality** | 0,8759 | 0,8428 | 0,8897 | 0,8826 | 0,9989 | 0,8263 | 0,8089 | 0,9773 | 0,9982 | 0,8279 |
| **Complexity** | 0,8445 | 0,8246 | 0,8942 | 0,9057 | 0,9993 | 0,8436 | 0,8352 | 0,9998 | 0,9991 | 0,8449 |

*Note.* All correlations are significant with p < .01.

**Table 2b.** Stability of structural and semantic variables in Network B.

|  | Removed Nodes | | | | | | | | | |
|---|---|---|---|---|---|---|---|---|---|---|
| **FORUM NETWORK C** (Pearson's Correlation Coefficients) | **Top 1st Percentile** | **Top 1st Perc. and Bottom** | **Top 5th Percentile** | **Top 10th Percentile** | **Bottom Nodes** | **Moderator Nodes** | **Moderators and Bottom Nodes** | **Spammer Nodes** | **Spammer and Bottom Nodes** | **Spammer and Mod. Nodes** |
| **Alter ART** | 0,9377 | 0,9377 | 0,9975 | 1,0000 | 0,9975 | 1,0000 | 1,0000 | 1,0000 | 0,9975 | 1,0000 |
| **Ego ART** | 0,9963 | 0,9963 | 0,9833 | 1,0000 | 1,0000 | 0,9994 | 0,9994 | 1,0000 | 1,0000 | 0,9994 |
| **Alter Nudges** | 1,0000 | 1,0000 | 1,0000 | 1,0000 | 0,9993 | 1,0000 | 1,0000 | 1,0000 | 0,9993 | 1,0000 |
| **Ego Nudges** | 0,9758 | 0,9758 | 1,0000 | 1,0000 | 1,0000 | 1,0000 | 1,0000 | 1,0000 | 1,0000 | 1,0000 |
| **Activity** | 0,9549 | 0,9533 | 0,7794 | 0,6183 | 0,9998 | 0,9866 | 0,9861 | 1,0000 | 0,9998 | 0,9866 |
| **Contribution Index** | 0,5437 | 0,4850 | 0,2913 | 0,2084 | 0,9902 | 0,5553 | 0,5159 | 0,9979 | 0,9878 | 0,5544 |
| **Betweenness Centrality** | 0,7637 | 0,7595 | 0,4078 | 0,1870 | 0,9999 | 0,9900 | 0,9896 | 1,0000 | 0,9999 | 0,9900 |
| **Betw. Cent. Oscillations** | 0,7402 | 0,7333 | 0,4071 | 0,3114 | 0,9960 | 0,8499 | 0,8445 | 0,9995 | 0,9958 | 0,8500 |
| **Closeness Centrality** | 0,2106 | 0,0541 | 0,0998 | 0,0586 | 0,9944 | 0,3119 | 0,2068 | 1,0000 | 0,9944 | 0,3119 |
| **Degree Centrality** | 0,9564 | 0,9602 | 0,8508 | 0,6866 | 1,0000 | 0,9982 | 0,9983 | 1,0000 | 1,0000 | 0,9982 |
| **Sentiment** | 0,8691 | 0,8622 | 0,9096 | 0,9089 | 0,9986 | 0,9198 | 0,9191 | 0,9988 | 0,9934 | 0,9201 |
| **Emotionality** | 0,8358 | 0,8268 | 0,8272 | 0,8844 | 0,9993 | 0,9217 | 0,9208 | 0,9992 | 0,9897 | 0,9230 |
| **Complexity** | 0,7979 | 0,8254 | 0,8397 | 0,7027 | 0,9974 | 0,8730 | 0,9982 | 0,9978 | 0,9982 | 0,8776 |

*Note.* All correlations are significant with $p < .01$.

**Table 2c.** Stability of structural and semantic variables in Network C.

Results show a very high degree of stability in all the node level metrics when removing peripheral nodes, spammers and the combination of the two classes of actors. This confirms the findings in Tables 1a-1c, i.e. the opportunity for an analyst to remove spammer and bottom nodes from a forum network, without greatly affecting the results of the analysis. By contrast the removal of moderators, or of overly-connected nodes, produced higher instability in the network metrics. In general, we notice that Ego/ Alter ART and Ego/Alter Nudges are rather stable metrics (except for Ego Nudges in Network B). Activity and Degree Centrality seem to be relatively stable, but they are progressively affected when the removal of overly connected nodes becomes significant, i.e. for removal of actors in the top $5^{th}$ and $10^{th}$ percentile of the degree distribution. Betweenness centrality and betweenness centrality oscillation are still relatively stable when removing moderators, even if betweennes centrality is affected more in Network A, when the network is in its early stages and moderators play a more important role. On the other hand, measures such as closeness centrality and contribution index seem to be more vulnerable to these node removal strategies.

Semantic variables show a trend which is similar to the other measures, as they are stable when removing spammers and bottom nodes, but they are progressively affected by the removal of actors with higher degrees. It is worth noting that the sentiment, complexity and emotionality of the language are less affected by the removal of moderators (who probably use a more "institutional" language), than by the removal of nodes in the top percentiles (which include non-moderators).

The results on the stability of node level metrics are consistent with those referred to the whole network variables. In other words, removing spammers and peripheral nodes can be a successful strategy to simplify the network and remove "noise" generated by these kinds of social actor, with no significant impact on the stability of structural and semantic variables. By contrast, removing moderators, as well as the deletion of the most connected nodes, can seriously harm the robustness of the network, altering the values of all the whole network metrics and of the majority of node level variables.

As the last step of our analysis, we looked for traits characterizing the behavior of moderators. We combined the results obtained from the three networks and implemented a t-test for each node-level variable, to study the differences in the behaviors of moderators. Results for significant differences are presented in Figure 1. Findings from this last analysis can be used to identify moderators and describe their behavior, when a list of moderators is not available from the forum managers.

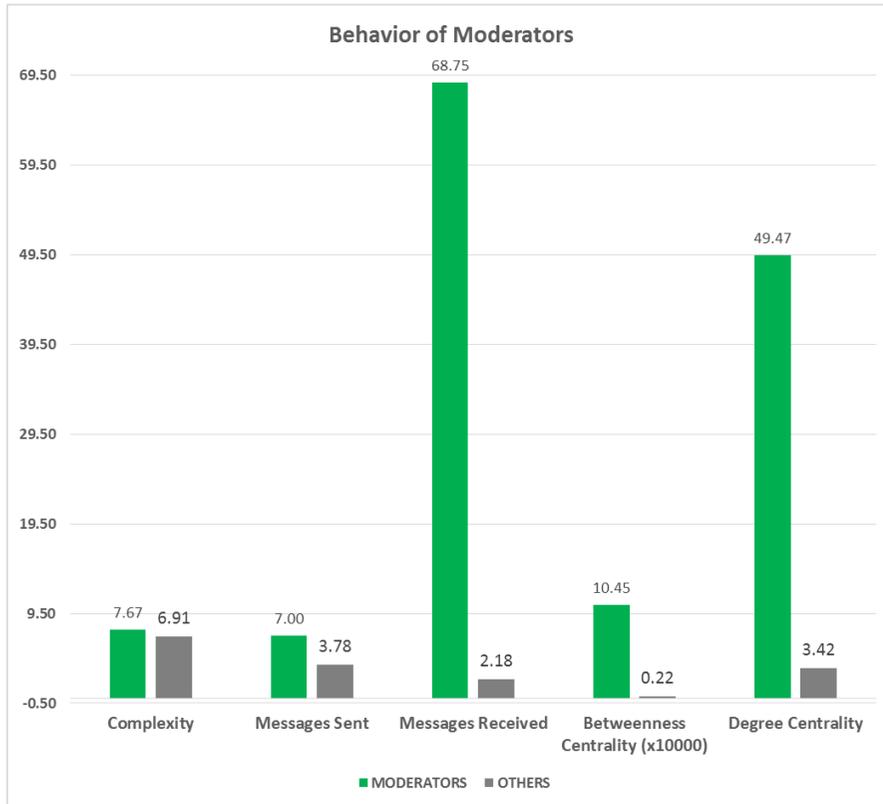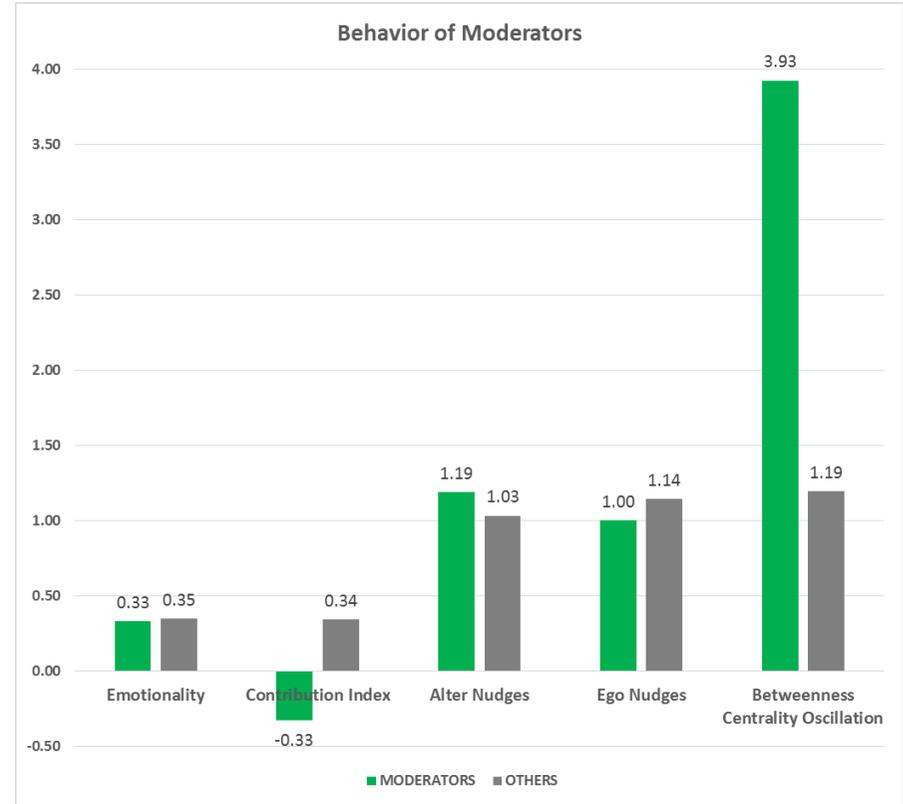

**Figure 1.** Behavior of moderators (significant t-tests, p < .05).

Moderators show much higher values of betweenness and degree centrality and betweenness centrality oscillations, probably due to their coordinating role. In all our forum networks employees often interacted with moderators with the purpose of obtaining more information about internal news and activities (such as the introduction of new badges, or changes in the work schedule). For this reason, moderators received and sent much more messages than rest of users and were pinged more by others to get an answer; they also received more messages than they sent – answering to multiple requests with a single post, thus having a negative contribution index. Moderators' answers, still high in number, were on average less emotional and more complex, as a consequence of the new knowledge they were introducing in the network and of their more controlled language. Differences in alter and ego nudges and in the language complexity and emotionality were less prominent, if compared with the other significant variables in the t-tests.

**5. Discussion, Practical Implications and Conclusions**

In this study, we explored the concept of network robustness, while applying several node removal strategies useful for social network analysts who are interested in evaluating the impacts of graph simplification techniques, and for community managers who want to better understand the role of forum moderators. Our research questions were inspired by the idea of identifying some specific sets of nodes which could be removed from intranet social networks without significantly harming their connectivity and while keeping structural and interaction metrics stable. In particular, we focused our attention on the effects produced when removing moderators, spammers, as well as network hubs or loosely connected nodes, located at the network periphery.

Our case study was carried out on three enterprise intranet forums, comprising more than 52,000 messages and about 12,200 user accounts, over a period of eight months.

We found that removing spammers and very peripheral nodes seems to be the best strategy to simplify the network and to get rid of "noise" generated by these types of social actor, without having a significant impact on structural and semantic variables. This finding can be particularly important if considered together with past studies proposing several techniques for the identification of spammers (Almeida & Yamakami, 2012; Heydari et al., 2015; Ramachandran & Feamster, 2006; Zheng et al., 2015) and if related to the possible negative effects of spam – such as producing distortions in network analyses, or working time losses for employees (Caliendo, Clement, Papies, & Scheel-Kopeinig, 2012). Spam messages also showed their negative influence in other fields, where it was proved that they can affect stock prices and investment choices in financial markets (Böhme & Holz, 2006; Hanke & Hauser, 2008).

Differently from the removal of spammers, the removal of moderators seems to significantly harm the network connectivity and the shared content (even if with a smaller impact). The most affected variables are closeness centrality and contribution index. We also found that the removal of overly connected nodes can significantly change the network structure, causing a deep distortion in almost all the measures considered both at ego and whole network level, especially when removing the nodes in the top $10^{th}$ percentile of the degree distribution. This is consistent with the previous studies conducted on scale-free networks (Albert et al., 2000; Gao et al., 2015). Accordingly, spammers should be identified and removed by criteria other than their degree centrality – such as looking at the content of their messages and at their contribution index, which should be close to 1. However, we also found measures which remained relatively stable, regardless of the removal strategy used: the average response time (Ego and Alter ART), the language sentiment, emotionality and complexity. These findings are relevant to those analysts and scholars who need to simplify a community graph before carrying out a more in-depth analysis. Graph simplification is sometimes beneficial in the reduction of the computational complexity, when resource intensive algorithms are applied, or when the presence of certain social actors (like spammers) can negatively affect other connected analyses

– such as a topic modeling of the forum content (Abainia, Ouamour, & Sayoud, 2016; Liang & de Rijke, 2016). Our results prove that once the spammers have been identified, they can be safely removed, together with the most peripheral nodes, without a big impact on the stability of structural and semantic variables; the same is not true for forum moderators. In general, a graph simplification which implies a reduction of the number of nodes can support the parameter estimation of well-known and widely used network models, such as the Exponential Random Graphs Models (Lusher, Koskinen, & Robins, 2013) or the stochastic actor-oriented models included in the R package RSiena (Ripley, Snijders, & Preciado, 2017): when likelihood-based computations are used, a too large graph size can make the parameter estimation infeasible (He & Zheng, 2013).

Lastly, we compared the behavior of moderators with the other users, with the aim of finding some distinctive characteristics through which moderators could be identified when their list is unknown. We found that moderators show a significantly higher betweennes centrality with also higher oscillations, a higher number of messages sent and received, degree centrality, alter nudges and complexity, and a lower emotionality, contribution index and ego nudges. Consistently with previous research (Berge & Collins, 2000; Brace-Govan, 2003), these findings indicate a role of people who are very central in the networks and who dynamically lead the forum discussions. Their language, often conveying the institutional message of the firm, is less emotional, but more complex (introducing new knowledge, when answering to questions). These last results we obtained are helpful for community managers who aim to better understand the role played by moderators and their impact on network connectivity. In general, we saw that moderators are important figures that if removed will significantly harm the interaction patterns among the other employees; therefore, they should be carefully selected by company managers for their ability to foster the online dialogue and social interactions, and for the new knowledge they could introduce to the networks. These findings support the idea that forum moderators often bring in their expertise about the forum topics and have a major role in shaping the online discussion and interactions (Frith, 2014; Huh et al., 2013). In addition, we maintain that it might be useful to assign the formal role of moderator to those employees who spontaneously behave in a similar way. Managers can refer to our results to identify these employees, looking for those who receive a high number of messages (who often correspond to people sharing information or advices), who are more central and with a higher number of direct connections, who answer quickly and oscillate more, thus supporting a rotating leadership process in one or more communities.

Limitations derive from our choice of examining few real-world networks, instead of using a simulated approach. This allowed a precise identification of moderators and spammers, but reduced the number of networks we could study. The number of spammers in our networks was rather low (less than 1%); accordingly, we advocate further research to better explore the role of these social actors, on datasets where their number is more significant. Moreover, we focused our attention on enterprise forum networks, but it would be interesting to analyze other kind of social platforms, such as other online forum communities or knowledge sharing websites (e.g. Stack Overflow). Lastly, we examined three real social networks with degree distributions similar to a power law: it would be interesting to conduct future research on larger datasets, comprising networks with different structures and from different business contexts. We encourage network analysts to repeat our experiment in their study contexts, to test the stability of the metrics they want to consider and to assess the set of nodes which can be safely removed before the analysis.

**Funding**

This research did not receive any specific grant from funding agencies in the public, commercial, or not-for-profit sectors.